\documentclass{iau} 
\usepackage{graphicx}
\usepackage{natbib}
\usepackage{color}

\title[Hierarchical information combination] 
{Learn from every mistake! \\ Hierarchical information combination in astronomy}

\author[M. S\"uveges et al.]   
{Maria S\"uveges$^1$\thanks{Present address: Max Planck Institute for Astronomy, K\"onigstuhl 17, 69117 Heidelberg, Germany.  Corresponding email: {\tt sueveges@mpia.de}},
Sotiria Fotopoulou$^1$,
Jean Coupon$^1$,
 St\'ephane Paltani$^1$,
 Laurent Eyer$^2$
 \and Lorenzo Rimoldini$^1$}

\affiliation{$^1$ Dept. of Astronomy, University of Geneva, Ch. d'Ecogia 16, 1290 Versoix, Switzerland \\[\affilskip]
$^2$ Dept. of Astronomy, University of Geneva, Ch. des Maillettes 51, 1290 Versoix, Switzerland}

\pubyear{2015}
\volume{325}  
\jname{Astroinformatics 2016}
\editors{M. Brescia, S.G. Djorgovski, E. Feigelson, G. Longo \\ \& S. Cavouti, eds.}
\begin{document}

\maketitle

\begin{abstract}
Throughout the processing and analysis of survey data, a ubiquitous issue nowadays is that we are spoilt for choice when we need to select a methodology for some of its steps. The alternative methods usually fail and excel in different data regions, and have various advantages and drawbacks, so a combination that unites the strengths of all while suppressing the weaknesses is desirable. We propose to use a two-level hierarchy of learners. Its first level consists of training and applying the possible base methods on the first part of a known set. At the second level, we feed the output probability distributions from all base methods to a second learner trained on the remaining known objects. Using classification of variable stars and photometric redshift estimation as examples, we show that the hierarchical combination is capable of achieving general improvement over averaging-type combination methods, correcting systematics present in all base methods, is easy to train and apply, and thus, it is a promising tool in the astronomical ``Big Data'' era.
\end{abstract}

\section{Introduction} \label{sec:intro}

In this era of massive surveys and resulting colossal databases, one of the hottest topics is how to mine relevant information from these data as efficiently as possible. Scientists are very inventive in constructing a wide diversity of methods to reach their goals and deriving scientific results in many ways. Very often, the issue the data analyst faces is not ``How to answer my question?'' but ``Which of the $n$ possible methods would solve best my problem?''. 

Two examples of this situation among many are the classification of astronomical objects and the estimation of photometric redshifts. Both subjects have recently seen the proliferation of methods used. For the first, there is nowadays an increasing variety of supervised classifiers; in the astronomical literature, \cite{dubathetal11,rimoldinietal12, goldsteinetal15, devineetal16, tramacereetal16} represent a few examples. The Gaia Variability Processing Pipeline \citep{DPACP-15} applies three methods for the classification of variable stars, Bayesian Networks, Gaussian Mixtures and Random Forest. Often, when many classifiers can be applied for the task, generally well-performing ones might fail on some classes while generally lower-performing others succeed on them, and it would be desirable to join the overall good performance of the former with the class-specific good scores of the latter. For photometric redshifts, many variants for both the empirical and the template-fitting methodologies exist \citep[e.g.][]{ilbertetal06,brammeretal08,carlilesetal10,carrascokindbrunner14b,hoyle16}, some of which (and more) are considered and tested for use in the Euclid Photometric Redshift pipeline. It would be useful to be able to merge the complementary advantages of these. 

We propose a high-dimensional meta-model to combine the output of several methods, which is generally applicable in every study where several models are possible to use, each of which gives a probability distribution of the parameter of interest as result. We use the above two specific topics to demonstrate the capabilities of this hierarchical setup, and show promising results on both of our examples. 

\section{Methodology} \label{sec:method}

Assume that $x_1, \ldots, x_K \in \mathcal{X}$ represents the data about an object of interest. This can be the multi-filter photometry of a galaxy, the attribute vector (amplitude, period, Fourier coefficients, colour, absolute magnitude) characterising a variable star, or any other informative set of measured values about a target. We wish to make inference about a parameter $\theta$ of the object; in the case of the galaxy, $\theta$ may be the redshift $z$, in the case of the variable star, its type. Suppose moreover that we have $M$ alternative methods, each of which yields a probability distribution function (PDF) of the parameter $\theta$: $p_1(\theta \mid x_1, \ldots, x_K), \ldots, p_M(\theta \mid x_1, \ldots, x_K)$. For photometric redshift estimation, these methods can be any variants of both the template fitting and the empirical methods, and for classification of variable stars, any supervised method from Gaussian mixtures to Random Forest, SVM or neural networks. The outputs $p_i(\theta \mid x_1, \ldots, x_K)$ are the probability distributions of the parameter of interest, conditioned on the observed data; point estimates of the parameter and uncertainties can be defined in multiple ways.

Our goal is to learn a new, better PDF $p_{\mathrm{comb}}(\theta | x_1, \ldots, x_K)$ of the true parameter value by combining these outputs. We look for a mapping $\mathcal{P} : p_1(\theta \mid x_1, \ldots, x_K), \ldots, p_M(\theta \mid x_1, \ldots, x_K) \longmapsto p_{\mathrm{comb}}(\theta | x_1, \ldots, x_K).$ The most commonly used way for this is weighted averaging of the outputs, often using different weights in different regions of $\mathcal{X}$ motivated by Bayesian Model Averaging (BMA), Bayesian Model Combination (BMC) or other \citep[e.g.][]{dahlenetal13, carrascokindbrunner14a}. Instead of this, we propose to learn this mapping by applying a nonparametric machine learning method, similar to Wolpert's proposition called ``stacking'' \citep{wolpert92}, which consists of a second-level, hierarchical training on the predictions from the base methods in a cross-validation-like setup. Suppose we have a training set, which we divide into two parts, one called $T$ (training) and the other $C$ (combination). 

\vspace{1mm}
\begin{enumerate}
\item[\bf{Step 1.}] We train the $M$ base methods on $T$.  We predict the PDFs on $C$, obtaining $M$ output PDFs, $ p_1(\theta \mid x_1, \ldots, x_K), \ldots, p_M(\theta \mid x_1, \ldots, x_K)$ for each of our objects in $C$. These can be represented each as the vector of values taken by the PDFs over a grid of $\theta$, or in some basis function system \cite[e.g.][]{carrascokindbrunner14c}.
\item[\bf{Step 2.}] In the next stage, we train a nonparametric learning method using the concatenated PDFs obtained on the objects of $C$ and the corresponding known values of the parameter of interest. The output of this second-level model is optimally a new probability distribution of the parameter. Thus, we learn the mapping leading to the final estimate $p_{\mathrm{comb}}(\theta | x_1, \ldots, x_K)$. Any machine-learning method can in principle be used here, though there may be technical restrictions on the choice, such as an ability to deal with high-dimensional data if its input PDFs have a high-dimensional representation. 
\item[\bf{Step 3.}] To predict a combined PDF of the parameter of interest of a new object, we first apply the trained base learners of Step 1 to obtain their output PDFs, then feed the concatenated PDFs into the trained combination model to obtain the combined result. 
\end{enumerate}
\vspace{1mm}

Optionally, the partition of the full training set into $T$ and $C$ can be repeated randomly $R$ times, Steps 1-3 can be performed using the ensemble of the $R$ models, and an average estimate of final combined PDFs may replace the single $p_{\mathrm{comb}}(\theta | x_1, \ldots, x_K)$. With such a procedure, as usually with ensemble methods, we can obtain more stable results, and can gather information about the uncertainty of the estimates due to training set selection. 

\begin{figure}[h]
\begin{center}
 \includegraphics[width=0.7\textwidth]{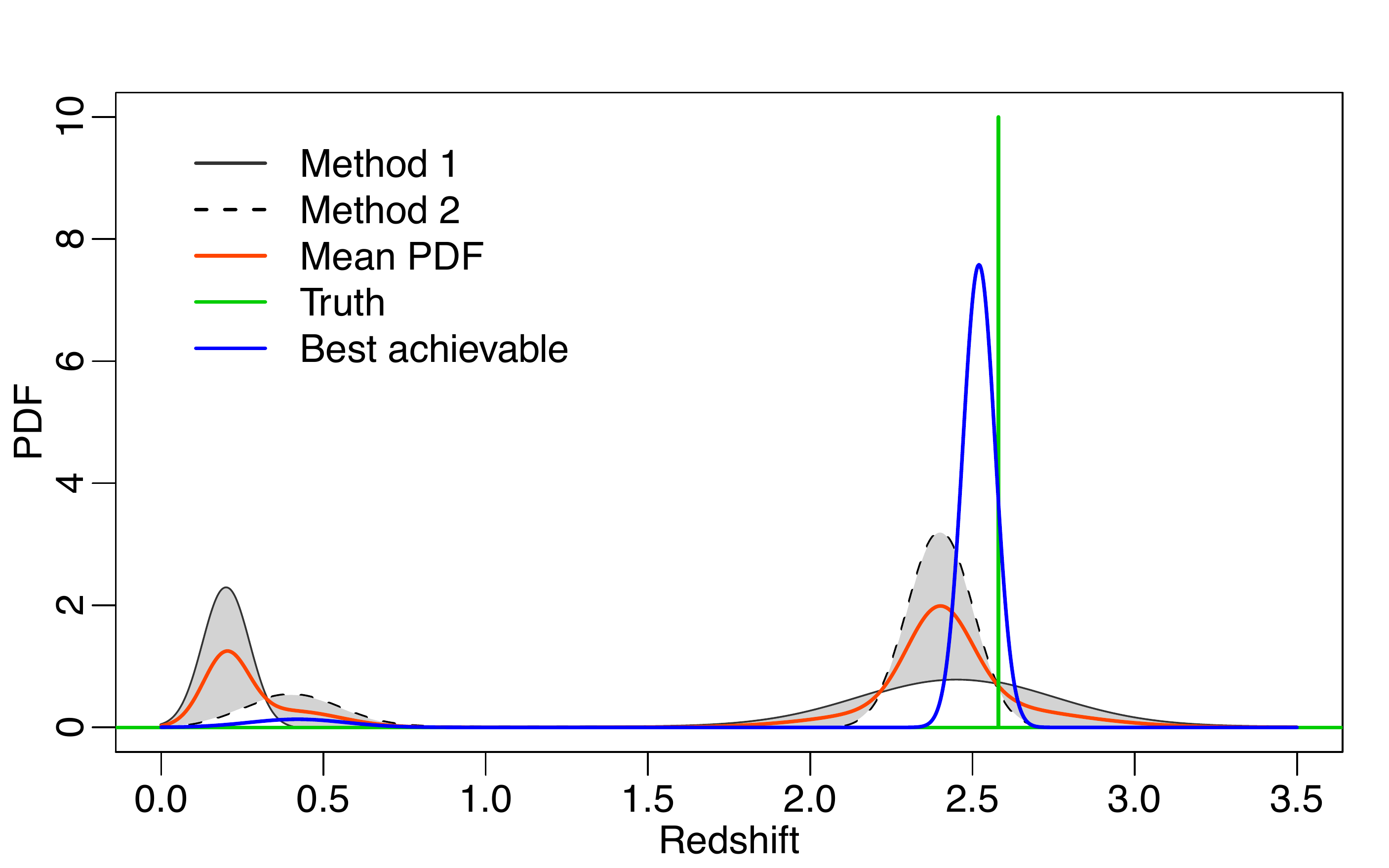} 
 \caption{An example of the results from two base methods (Method~1 and 2, black solid and dashed), the average (red), the optimal PDF obtainable with the data (blue), and the true value (green).}
   \label{avgLimitation}
\end{center}
\end{figure}

Why can this hierarchical learning be expected to work better than BMA or BMC? Figure~\ref{avgLimitation} depicts the main reason. Suppose that we have two base methods (Method~1 and 2) that yield the probability distributions for the photometric redshift of a test galaxy, shown in black solid and black dashed lines in the figure. The true redshift of the galaxy, shown as a green spike, may be known from spectroscopic measurements, and indicates a bias for Method~2 and a catastrophic error for Method~1. Both of these mistakes can be characteristic of the method used (due perhaps to the chosen template set for the template-fitting Method~2 and the dominance of low-redshift objects in the training sample for the empirical Method~1), and not an implication of the data: with an ``optimal'' method (better-adapted templates for Method~2, a balanced sample for Method~1 or a third method), we might be able to reduce the bias and obtain the ``best achievable'' estimate with our data (in blue in the figure). Combinations based on averaging can never leave the grey area delimited by the two base PDFs, and cannot approximate well the ``best achievable'' PDF. This remains so using any number of methods, or partitioning of the input space $\mathcal{X}$ into several regions: at each $x_1, \ldots, x_K \in \mathcal{X}$ the combined PDF will remain between the upper and lower envelope of the PDFs from the methods. In other words, systematics common to the methods cannot be corrected by using a linear combination. The nonparametric, unconstrained learning proposed above is in principle able to learn also nonlinear relationships, and thus approximate the nonlinear mapping from Method~1 and 2 to the ``best achievable''.

Moreover, the outputs of the base methods are not independent. They use often the same or overlapping data, and in many cases, are built on similar principles with possibly only small differences; an example for this is the template-fitting photometric redshift estimating methods using principal component decomposition or different pre-determined sets of templates. Combining two such methods requires accounting for their dependency as well. This dependence must be modelled when we seek the estimating mapping $p_1(\theta \mid x_1, \ldots, x_K), \ldots, p_M(\theta \mid x_1, \ldots, x_K) \longmapsto p(\theta | x_1, \ldots, x_K)$. Averaging does not take into account this dependence structure, while the above proposed method is able to learn and thus potentially make use of it. In other words, it may be able to ``learn from the mistakes of all''.

\section{Data and applications}

We tested our procedure using the following general framework. We selected a known, thoroughly analysed dataset from the literature for both variable star classification and photometric redshift estimation, and drew random partitions over them into three equal parts ($T$ for base training, $C$ for combination, $V$ for validation of the results) $R$ times (for variable stars, $R = 1000$, for photometric redshifts, $R = 500$). The procedure described in Section \ref{sec:method} was run on each partition, using a Random Forest learner \citep{breiman01} for the combination because of its stability, simplicity, insensitivity to tuning parameters and ability to deal with high-dimensional data. For a fair comparison, we trained the base methods also on the joined $T\cup C$ set, thus providing them the same amount of training data as to the combination. We compare the combinations to these ``doubly-trained'' base classifiers, in order to ensure that the improvement by the combination is not simply due to the twice as large training set. We also computed a combination based on weighted averages, where the weights were taken to be proportional of the fraction of correct predictions by the method on set $C$. The results presented are averaged over all sets $V$.

\subsection{Classification of variable stars}

The used complete dataset consists of 1661 stars from the Hipparcos periodic variable catalog \citep{perrymanetal97, esahipparcos}, and used in \citet{dubathetal11}. The class system was simplified to 15 classes, merging some of them (such as all subtypes of classical Cepheids into CEPCL or Type-II Cepheids into CEPT2) and omitting very rare classes, obtaining finally classes that had each at least 20 members. The objects were characterised by attributes derived from their light curve (period, amplitude, Fourier amplitudes and phases, statistical summaries such as skewness and kurtosis) complemented by visual and near-infrared colours \cite[for details, see][]{dubathetal11}. We used five base classifiers: C5.0, Random Forest (RF), Gaussian Mixtures (GM), Support Vector Machines (SVM) and Linear Discriminant Analysis (LDA) \citep[a textbook summarizing them is][]{hastieetal}. 

\begin{figure}[h]
\begin{center}
 \includegraphics[width=0.45\textwidth]{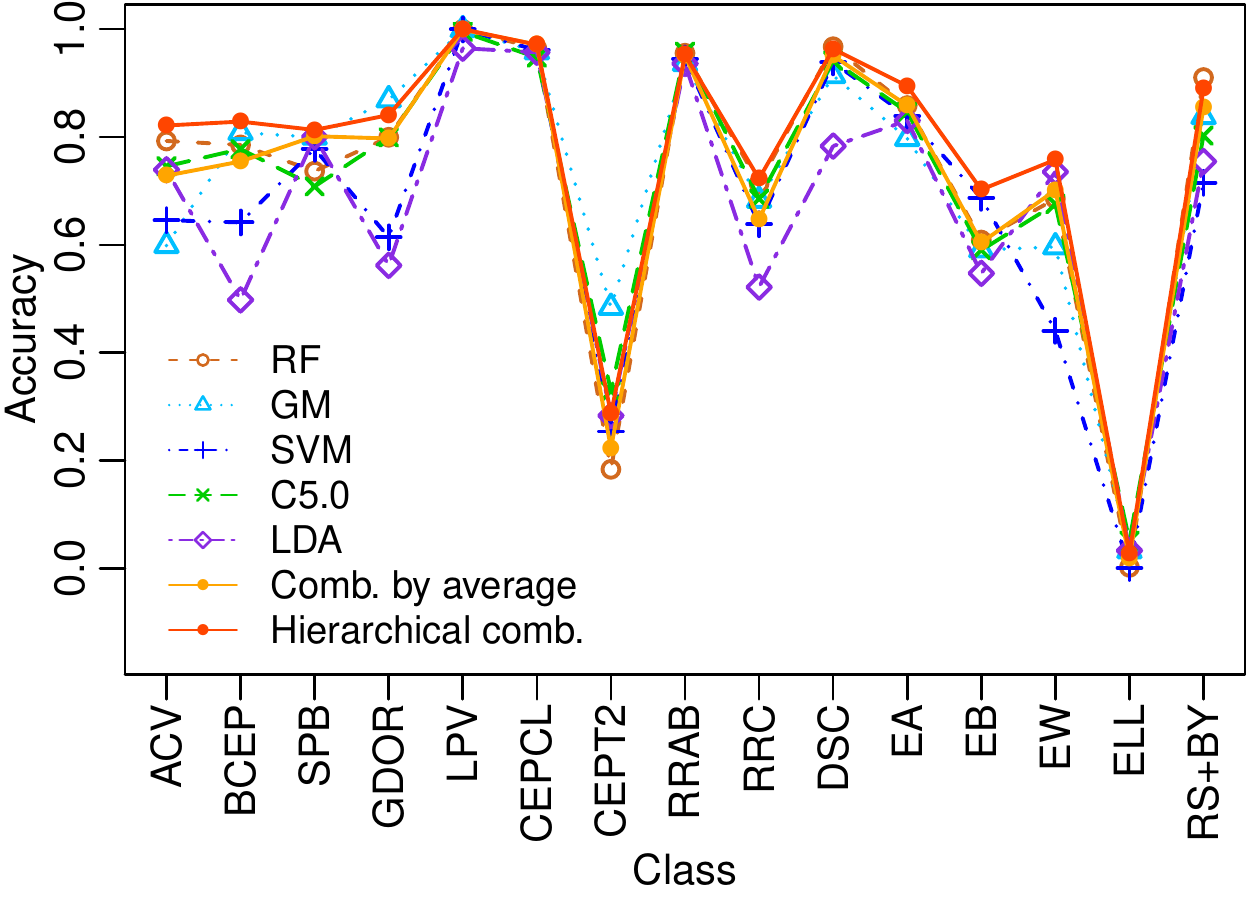} 
 \hspace{2mm}
 \includegraphics[width=0.52
 \textwidth]{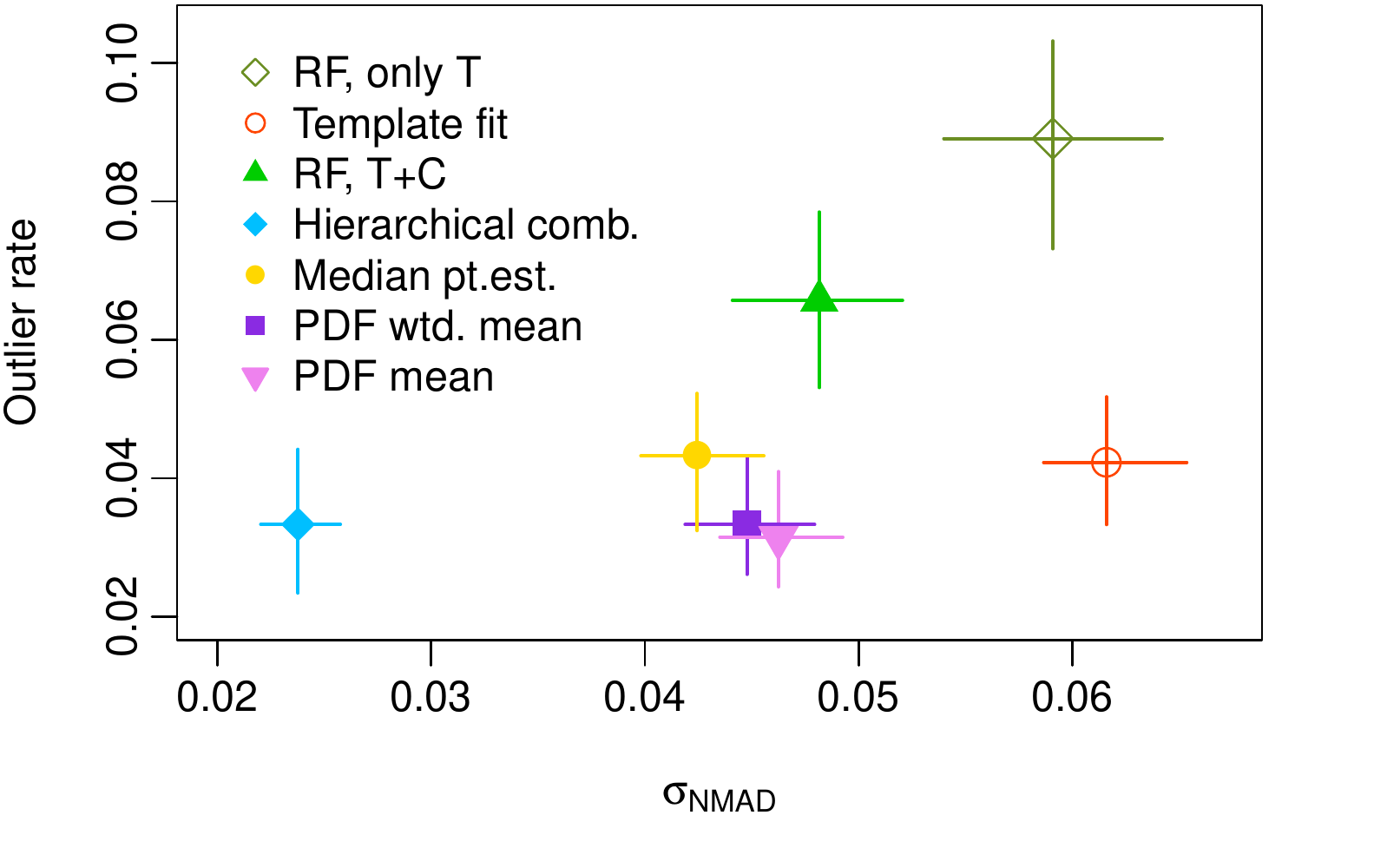} 
 \caption{The performance of the base learners compared to variants of the probabilistic information combination for  variable star classification (left panel) and for photometric redshift estimation (right panel). Dashed and dash-dotted lines with superposed empty symbols in the left panel show the classwise accuracies (the fraction of correctly estimated class labels) of the five base methods, averaged over the validation sets of 1000 random partitions of the Hipparcos data. The solid lines combined with filled symbols present the performance of an average-based combination (orange) and the hierarchical combination (red). In the right panel, the fraction of catastrophic outliers is shown against the normalized median absolute deviation for the estimation of the photometric redshifts (for the definitions, see text). The bars represent the (0.025,0.975) quantiles of the values obtained on the 500 random partitions.}
   \label{fig2}
\end{center}
\end{figure}

The classwise performances of the base classifiers (trained on $T\cup C$) are shown with dashed and dash-dotted lines and empty symbols in the left panel of Figure~\ref{fig2}. In particular, the overall weakest performance of LDA and its point of excellence, the classification of the class EW (contact eclipsing binaries) are visible. An ideal combination should preserve this excellence of LDA in identifying EW objects, while on the other classes, maintain the overall good performance of the other classifiers. The hierarchical combination, shown in red solid line, is very close to achieve this: either it exceeds all the base classifiers, or is very near to the maximum accuracy, including on the EW class. The combination by average in comparison is somewhat weaker on several classes. The mean global accuracies are 82.5\% (RF), 81\% (C5.0), 79.9\% (GM and SVM), 77.9\% (LDA) for the base classifiers, while they are 82.2\% for the averaging combination and 85.8\% for the hierarchical combination. This shows an average improvement of 3.3\% by the latter over the best base method. The detailed results on the 1000 random partitions show that it yields improvement over the best base classifier in all partitions. 
 
\subsection{Photometric redshift estimation}

The data, containing $u,g,r,i,z,J,H$ and $K_s$ photometry of 3331 galaxies, are a subset from field D1 of the WIRDS-CHFTLS database \citep{bielbyetal12}, having spectroscopic redshifts from the VIMOS-VLT Deep and UltraDeep surveys \citep{lefevreetal05, lefevreetal13}. To estimate the redshift of the objects, we implemented a least squares template fitting algorithm without photometric zero-point  calibration, using the COSMOS templates with no template calibration or added emission lines, and an empirical RF regression, using only the default tuning parameters proposed by \cite{breiman01}. We trained these base methods (both on only $T$ and on $T\cup C$), the hierarchical and several average-based combinations\footnote{These included a weighted and a non-weighted mean of the base PDFs trained on $T$, and the median of the two point estimates from the base methods. The weights for the first were defined to be proportional to the fraction of catastrophic outliers on $C$ (see the definition later).} on the 500 random partitions. To produce the presented plots, we computed several point estimates of $z$ from the PDFs provided by the base methods and by the combinations for all objects when they were in set $V$, and selected the best of these (mean of the PDF for the template fit, median of the PDF for the base RF and the combinations). In what follows, this best point estimate is denoted by $z_{\mathrm{ph}}$, and the spectroscopic redshift which is considered to be the truth by $z_{\mathrm{sp}}$.

The right panel of Figure~\ref{fig2} shows the catastrophic outlier rate (the fraction of objects for which $| z_{\mathrm{sp}} - z_{\mathrm{ph}}| / (1 + z_{\mathrm{sp}}) > 0.15$) against the normalized median absolute deviation (defined as $\sigma_{\mathrm{NMAD}} = 1.48\times \mathrm{median}| z_{\mathrm{sp}} - z_{\mathrm{ph}}| / (1 + z_{\mathrm{sp}})$). The best methods fall therefore at the lower left corner of the plot. The base methods have by far the largest scatter, and the base RF models (both trained on $T$ and $T\cup C$) the largest outlier fraction.  The improvement that can be obtained by using a larger training set with the same single method can be seen when comparing the two RF models: both the scatter and the outlier fraction decreases notably. Not so much, however, as obtained by variants of the averaging combination (pink triangle, violet square and yellow dot). The overall best results, however, are furnished by the hierarchical combination, with a further strong decrease of the scatter.  It is remarkable that even though the second-level method in the hierarchical combination is an empirical machine-learning one, and therefore just as hampered by the underrepresentation of high-$z$ objects as the base RF, it still improves even on the outlier fraction of the template fitting base method.   

\begin{figure}[h]
\begin{center}
 \includegraphics[width=0.99\textwidth]{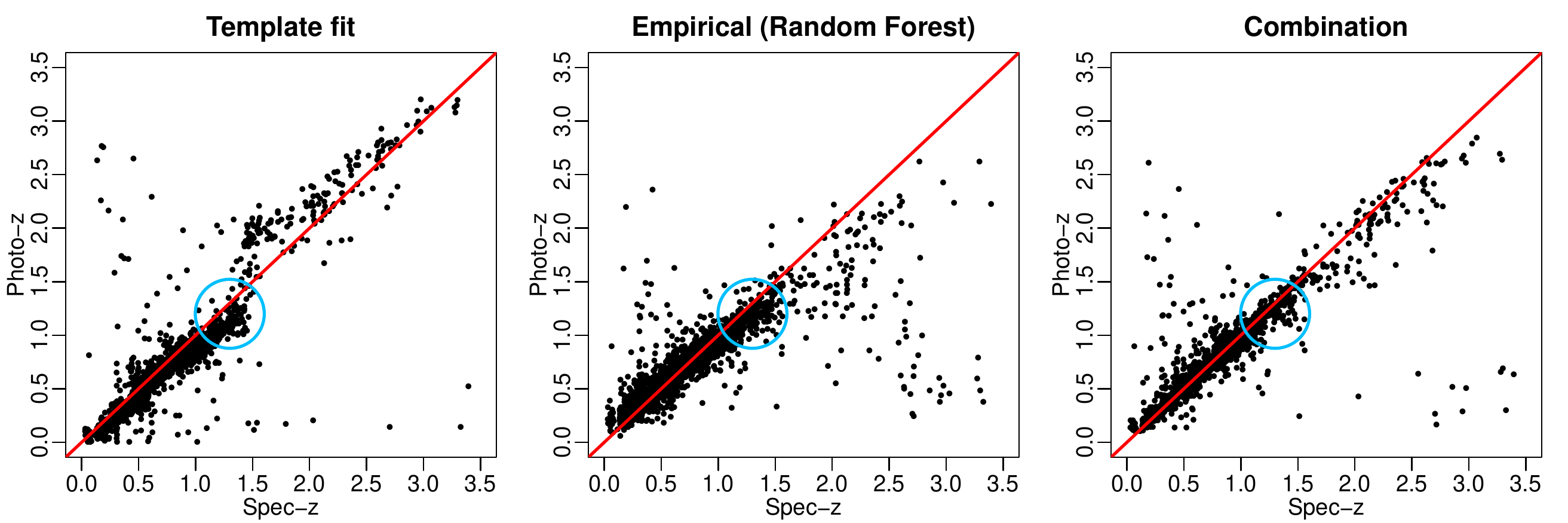} 
 \caption{Visualization of the results of hierarchical combination through the best point estimates versus the spectroscopic redshifts. Left panel: template fitting method, with the mean of the PDFs as $z_{\mathrm{ph}}$. Middle panel: empirical method, with the median of the PDFs as $z_{\mathrm{ph}}$. Right panel: hierarchical combination, with the median of the PDFs as $z_{\mathrm{ph}}$. The red line corresponds to  $z_{\mathrm{ph}} = z_{\mathrm{sp}}$. The blue circle emphasizes the neighbourhood where both base methods are biased downwards, which is visibly decreased by the combination.}
   \label{fig3}
\end{center}
\end{figure}

Figure~\ref{fig3} shows the $z$-dependent systematic bias of the template fits (left panel) and the high scatter and outlier rate of the empirical method (middle panel), when compared to the ideal $z_{\mathrm{ph}}=z_{\mathrm{sp}}$ line (red). The hierarchical combination (right panel) corrects the systematic biases of the first, and shrinks the scatter and decreases the number of catastrophic outliers in the high-$z$ regime of the second. Thus, it is indeed able to pick the best of both method, while learning to ignore their systematic failures. Moreover, it is able to do what an averaging method cannot: around redshifts 1.2-1.5, where both base methods are downward biased (blue circle), it largely removes this bias. The learning here is based on simultaneous presence of specific patterns in the output of the two methods, not on a straightforward pointwise averaging.

\section{Conclusions}

Our paper presents a general hierarchical information combination method, which is aimed at the efficient extraction of useful information from the data. The first level of the hierarchy trains several base methods (classifiers, regression models or any other statistical or machine-learning model) producing each a probability distribution of the parameter of interest. The second level of the hierarchy consists of training a second nonparametric machine-learning model (e.g. Random Forest) on the outputs of the base models. The  experiments on variable star classification and photometric redshift estimation show the following:
\begin{itemize}
\item The information in a given training set about a parameter is more efficiently exploited if the set is divided into two parts in order to train a hierarchical combination model using several base models than to train a single model with the complete training set. 
\item In both of our examples, we achieved always improvement in the global results over the best single model.
\item The hierarchical combination is able also to correct systematics and biases where {\it all} the base models are similarly biased.
\item The hierarchical combination is very general: it can be applied for every study where there are alternative methods providing different views of the data, producing probability distributions as output.
\item The choice of the combiner, though it must be able to model nonparametric relationships and high-dimensional data, is largely free. In our study, we used Random Forest, which in addition was little sensitive to tuning parameters.
\end{itemize}
In conclusion, our study shows on two examples that accepting diversity and unifying its various strengths into a synthesis appears the best strategy -- certainly in astronomical data analysis.


\end{document}